\documentclass[11pt]{article} %
\makeatletter

\newif\ifuserevtex
\newif\ifusetwocolfigs

\@ifclassloaded{revtex4-2}{%
  \userevtextrue
  \@ifclasswith{revtex4-2}{reprint}{%
    \usetwocolfigstrue
  }{%
    \usetwocolfigsfalse
  }%
}{%
  \userevtexfalse
  \usetwocolfigsfalse %
}

\makeatother

\newif\ifshowlinenumbers
\usepackage[T1]{fontenc}         %
\usepackage[utf8]{inputenc}      %
\usepackage[english]{babel}      %

\usepackage{amssymb}             %
\usepackage{mathtools}           %
\usepackage{dcolumn}             %

\usepackage{graphicx}            %
\graphicspath{{./}}

\usepackage{color,xcolor}        %
\usepackage{soul}                %
\usepackage{tcolorbox}           %
\usepackage{cancel}              %
\usepackage{comment}             %
\usepackage{moreverb}            %
\usepackage[normalem]{ulem}      %

\ifuserevtex
    \newcommand{\mybibliostyle}{apsrev4-2}
\else
    \usepackage{amsthm}
    
    \usepackage[a4paper,margin=2.3cm]{geometry} %
    \usepackage{appendix} %
    
    \usepackage[numbers]{natbib} %
    \newcommand{\mybibliostyle}{unsrtnat} %

    \usepackage{authblk} 
    \usepackage{setspace} %
\fi

\usepackage{pifont}

\usepackage{etoolbox}  %
\usepackage{environ}   %
\usepackage{xparse}
\usepackage{scalerel}

\usepackage[
  colorlinks=true,
  linkcolor=blue,  %
  citecolor=blue,  %
  urlcolor=blue    %
]{hyperref}

\usepackage[capitalise]{cleveref}  %

\usepackage{physics}  %

\newif\ifshowallemails
\showallemailsfalse  %

\makeatletter

\def\affillist{}
\newcommand{\myaffil}[2]{%
  \expandafter\def\csname affil@#1\endcsname{#2}%
  \ifuserevtex\else
    \listgadd{\affillist}{#1}%
  \fi
}

\newcommand{\buildaffiliations}{%
  \ifuserevtex\else
    \forlistloop{\emitoneaffil}{\affillist}
  \fi
}
\newcommand{\emitoneaffil}[1]{%
  \expandafter\affil\expandafter[#1]{\csname affil@#1\endcsname}%
}

\newcommand{\orcidlink}[1]{%
  \href{https://orcid.org/#1}{\textsuperscript{\textcolor{green!50!black}{\textbullet}}}%
}

\NewDocumentCommand{\myauthor}{O{} O{} O{} m m}{%
  \ifuserevtex
    \author{#4%
      \ifx&#2&\else\ \orcidlink{#2}\fi
    }%
    \ifx&#1&\else
      \ifshowallemails
        \email{#1}%
      \else
        \ifx&#3&\else  %
          \email{#1}%
        \fi
      \fi
    \fi
    \@for\aid:=#5\do{%
      \expandafter\affiliation\expandafter{\csname affil@\aid\endcsname}%
    }%
  \else
    \author[#5]{#4%
      \ifx&#2&\else\ \orcidlink{#2}\fi
      \ifx&#1&\else
        \ifx&#3&%
          \ifshowallemails
            \thanks{\href{mailto:#1}{#1}}%
          \fi
        \else
          \thanks{Corresponding author: \href{mailto:#1}{#1}}%
        \fi
      \fi
    }%
  \fi
}

\newcommand{\storedabstract}{}  %
\newcommand{\storedkeywords}{}  %

\newcommand{\myabstract}[1]{%
  \long\def\storedabstract{#1}%
}

\newcommand{\mykeywords}[1]{%
  \def\storedkeywords{#1}%
}

\newcommand{\maketitleandabstract}{%
  \ifuserevtex
    \begin{abstract}
    \storedabstract
    \end{abstract}
    \keywords{\storedkeywords}
    \maketitle
  \else
    \maketitle
    \section*{Abstract}
    \storedabstract
    \ifx\storedkeywords\@empty
    \else
      \vspace{1em}
      \par\noindent \textbf{Keywords:} \storedkeywords
    \fi
  \fi
}

\newcommand{\storedhighlights}{}  %

\newcommand{\myhighlights}[1]{%
  \ifuserevtex
  \else
    \gdef\storedhighlights{#1}%
  \fi
}

\newcommand{\inserthighlights}{%
  \ifuserevtex
  \else
    \par\noindent\textbf{Highlights:}
    \begin{itemize}
      \storedhighlights
    \end{itemize}
    \newpage
  \fi
}

\newcommand{\storedacknowledgements}{}  %

\newcommand{\myacknowledgements}[1]{%
  \gdef\storedacknowledgements{#1}%
}

\newcommand{\insertacknowledgements}{%
  \ifx\storedacknowledgements\@empty
  \else
    \ifuserevtex
      \begin{acknowledgments}
      \storedacknowledgements
      \end{acknowledgments}
    \else
      \section*{Acknowledgements}
      \storedacknowledgements
    \fi
  \fi
}

\makeatother

\ifshowlinenumbers
  \usepackage{lineno}
  \AtBeginDocument{%
    \ifuserevtex
      \ifusetwocolfigs
      \else
        \linenumbers
      \fi
    \else
      \linenumbers
    \fi
  }
\fi

\makeatletter

\def\onecolfig{\@ifnextchar[{\onecolfig@opt}{\onecolfig@opt[]}}
\def\endonecolfig{\end{figure}}
\def\onecolfig@opt[#1]{\begin{figure}[#1]}

\def\twocolfig{\@ifnextchar[{\twocolfig@opt}{\twocolfig@opt[]}}
\def\endtwocolfig{\ifusetwocolfigs \end{figure*} \else \end{figure} \fi}
\def\twocolfig@opt[#1]{%
  \ifusetwocolfigs
    \begin{figure*}[#1]%
  \else
    \begin{figure}[#1]%
  \fi
}

\makeatother

\usepackage{xr}

\makeatletter
\newcommand*{\addFileDependency}[1]{%
  \typeout{(#1)}
  \@addtofilelist{#1}
  \IfFileExists{#1}{}{\typeout{No file #1.}}
}
\makeatother

\newcommand{\papertitle}{Crack-Tip Opening as a Probe for Length-Scale Separation in Geometrically Nonlinear Solids
}

\newcommand{\nd}{\mathrm{d}} %
\newcommand{\pd}{\partial} %
\newcommand{\Det}{\mbox{Det}\,} %

\newcommand{\id}[1][ij]{\delta_{#1}} %
\newcommand{\gradU}[1][i,j]{u_{#1}}    %
\newcommand{\defGrad}[1][]{F_{#1}}   %
\newcommand{\greenL}[1][ij]{E_{#1}} %
\newcommand{\rightCG}[1][]{C_{#1}} %
\newcommand{\firstPK}[1][]{P_{#1}} %
\newcommand{\refT}[1][]{T_{#1}} %
\newcommand{\secondPK}[1][ij]{S_{#1}} %
\newcommand{\remoteLoad}[1][]{\sigma_{\infty}} %

\newcommand{\energyDen}{{\cal W}}  %
\newcommand{\Jint}{{\cal J}}
\newcommand{\err}{\cal G} %
\newcommand{\fractEn}{\Gamma} %
\newcommand{\jacDet}{J} %
\newcommand{\firstI}{I} %
\newcommand{\E}{E_\mathrm{m}} %

\newcommand{\refDomain}{{\Omega}}
\newcommand{\defDomain}{{\Omega^*}}

\newcommand{\bound}{{\partial \Omega}}
\newcommand{\closeC}{{\cal C}}

\newcommand{\angSoll}[1][]{U_{#1}}
\newcommand{\angOne}{M} %
\newcommand{\angTwo}{N} %

\newcommand{\lengthSCd}{\lambda_\mathrm{nl}} %
\newcommand{\lengthSCr}{\lambda_\mathrm{nl}^\mathrm{r}} %

\newcommand{\innersold}{\delta^\text{inner}}
\newcommand{\outersold}{\delta^\text{outer}}
\newcommand{\innersolr}{\delta^\text{inner}_\mathrm{r}}
\newcommand{\outersolr}{\delta^\text{outer}_\mathrm{r}}
\newcommand{\ctodd}{\delta}
\newcommand{\ctodr}{\delta_\mathrm{r}}

\newcommand{\ie}{{\textit{i.e.}}}
\newcommand{\beq}{\begin{equation}}
\newcommand{\eeq}{\end{equation}}
\newcommand{\bea}{\begin{eqnarray}}
\newcommand{\eea}{\end{eqnarray}}

\begin{document}

\title{\papertitle}
\myaffil{1}{Institute for Building Materials, ETH Zurich, Switzerland}
\myaffil{2}{Laboratoire de Physique, CNRS, ENS de Lyon, Université de Lyon, 69342 Lyon, France}
\myaffil{3}{Departamento de Física, Universidad de Chile, Casilla 487-3, Santiago, Chile}

\myauthor[][0009-0000-9226-6321]{Raúl Lazo-Molina}{1}%
\myauthor[][0000-0003-0584-8046]{Mokhtar Adda-Bedia}{2}%
\myauthor[][0000-0001-7244-7416]{Mohit Pundir}{1}%
\myauthor[][0000-0001-7627-2860]{Rodrigo Arias}{3}%

\myauthor[dkammer@ethz.ch][0000-0003-3782-9368][yes]{David S. Kammer}{1}

\buildaffiliations \date{\today}

\myabstract{
Soft elastic solids are highly deformable materials where fracture is driven by the complex coupling of geometric and material nonlinearities. While geometric nonlinearity (GNL) arises kinematically from the intrinsic capacity of solids to undergo large deformations, material nonlinearity stems from the constitutive behavior unique to each class of materials. Because GNL is a universal feature of all highly deformable solids, establishing its standalone impact is a prerequisite for understanding nonlinear fracture. Here, we focus on brittle soft solids to study the role of GNL alone on the near-tip fields of a static crack under mode I plane-strain conditions, providing a canonical baseline for integrating material nonlinearities in future investigations. By utilizing a compressible St.~Venant-Kirchhoff material model, we analyze crack behavior under large deformations in the absence of material nonlinearity. We propose a robust postprocessing methodology based on the crack-tip opening displacement (CTOD) profile and derive asymptotic analytical solutions. Our results reveal a distinct near-tip region where the CTOD departs from classical linear elastic predictions, transitioning into a nonlinear regime dictated by Poisson's ratio. Using a matched-asymptotics approach, we define a physical nonlinear length scale $\lengthSCd$ that bounds this region and scales quadratically with the far-field stress intensity factor $K_I$. We show that GNL acts as an~intrinsic strain-stiffening mechanism sufficient to trigger energy partitioning, effectively shielding the crack tip and imparting an apparent toughening. Ultimately, we conclude that the geometrically nonlinear material model serves as a foundational framework for the broader study of nonlinear elastic fracture mechanics.
}

\mykeywords{fracture mechanics, nonlinear elasticity, geometric nonlinearity, soft materials, asymptotic analysis, NLEFM, LEFM.}

\myhighlights{
    \item CTOD-based methodology to define its asymptotic solution and nonlinear length scale.
    \item Analytic solution for crack tip behavior and nonlinear length scale in SVK material.
    \item GNL model as foundation for understanding the effects of material nonlinearity.
}

\maketitleandabstract
\vspace{1cm}
\inserthighlights

\section{Introduction}

Soft materials such as elastomers, gels, and biological tissues have attracted significant research attention due to their capacity to sustain large reversible deformations and their compatibility with biological systems \cite{Long2021}. These properties make soft materials highly attractive across a broad range of applications, including biomedical load-bearing implants \cite{Li2024, Andrade2022}, soft robotics \cite{elango2015}, adhesives \cite{creton2016a}, and drug delivery systems \cite{Suksaeree2014}. Despite their high stretchability, many soft materials are inherently brittle, exhibiting low resistance to crack propagation and a tendency to fail catastrophically once a defect is introduced. This brittle behavior, which limits the adoption of soft materials in many engineering applications, is observed across a wide range of systems, including silicon elastomers such as PDMS \cite{Gonzalez2020}, single-network hydrogels like PAAm \cite{Kundu2009}, PEG-DA \cite{Yang2024}, and agar \cite{Barrangou2006}, and biopolymers, such as fibrin clots \cite{Liu2021b} and collagen fibers \cite{Gentleman2003}. Hence, to expand the range of possible applications of soft materials, it is necessary to enhance their fracture resistance, which requires a fundamental understanding of their failure mechanics.
The failure of brittle materials has classically been understood through the framework of linear elastic fracture mechanics (LEFM)~\cite{Anderson2017, Broberg1999}. LEFM relies on the assumptions of infinitesimal strains and small-scale yielding; this behavior results in a square-root singularity of the elastic fields near the crack tip and dissipative processes being confined to a point-like process zone. While these assumptions are valid for hard materials, such as glass and ceramics, they are commonly violated in soft materials. Indeed, experimental studies on numerous soft materials, including hydrogels~\cite{Livne2008, Livne2010, Pan2023}, gels~\cite{Seitz2009}, and silicone polymers~\cite{Ahmad2019, Qi2019}, have demonstrated that the asymptotic elastic fields often exhibit power-law singularities that differ from the square-root prediction of LEFM.  These singularities have also been found to be material-dependent and can exhibit anisotropy~\cite{Buehler2003, Wang2023a, Pundir2024}.  The precise nature of the singularity at the crack tip fundamentally dictates the physics of the fracture process, and the violation of the assumptions of LEFM can lead to novel behavior. For example, in the case of dynamic mode I fracture, although LEFM predicts the Rayleigh wave speed, $c_R$, to be the upper bound for crack propagation speeds, numerical simulations \cite{Buehler2003, Guozden2009, Marder2005, Marder2006, Pundir2024} and experiments \cite{Moulinet2015, Petersan2004, Mai2020, Wang2023a} on soft materials have observed crack velocities that not only exceed $c_R$ but even surpass the material shear wave speed, $c_S$, reaching the supershear range. This evidence underscores the importance of a more complete understanding of the failure of soft brittle materials.
Unlike hard materials, soft brittle solids exhibit finite strains over an extended region around the crack tip, which introduces significant nonlinear behavior that is not accounted for in LEFM. One notable theoretical approach to understanding this behavior is the weakly nonlinear theory of dynamic fracture~\cite{Bouchbinder2008}. This theory introduces a  nonlinear length scale that explains some of the discrepancies between LEFM predictions and observations from hydrogel experiments \cite{Livne2008}; however, as a perturbative approach, it breaks down near the crack tip where the strain fields exhibit extreme values \cite{Livne2010, GoldmanBoue2015a}.  To accurately describe the strain fields in this region, frameworks based on Knowles and Sternberg's asymptotic theory \cite{Knowles1973, Knowles1974} have examined static cracks using a variety of hyperelastic laws, including the neo-Hookean, Gent, exponential hardening, and generalized neo-Hookean material models \cite{Long2015}. All of these frameworks, however, directly incorporate both material nonlinearity, described through different constitutive laws, and geometric nonlinearity (GNL), arising from the common kinematic description of finite strains \cite{Holzapfel2000}. Since the effects of these two nonlinearities are treated together in existing studies, it is not straightforward to attribute a given phenomenon to a specific source of nonlinearity, which represents a significant limit in our understanding of crack behavior in soft solids. The importance of GNL was recently demonstrated when it was shown that GNL alone, even in the absence of material nonlinearity, is sufficient to enable dynamic oscillatory instability~\cite{Lubomirsky2018} and supershear crack propagation~\cite{Pundir2024}. Despite the clear importance of GNL, the underlying static crack-tip solution accounting solely for the effects of GNL remains unknown.
Here, we quantify the effects of GNL alone on the near-tip fields of a pre-existing crack under mode I loading. To this end, we numerically solve the static crack problem considering a compressible St.~Venant-Kirchhoff (SVK) material under plane-strain conditions. Because the SVK model combines a linear elastic constitutive relationship with nonlinear kinematics, it effectively enables the study of crack behavior in a soft elastic solid focusing on the effects of GNL in the absence of material nonlinearity. To systematically characterize the crack-tip asymptotic behavior, we introduce a postprocessing methodology based on the crack-tip opening displacement (CTOD) profile. The CTOD profile offers a direct advantage over strain- or stress-based measurements because it can be directly measured in experiments, thus avoiding the additional techniques necessary to obtain strain or stress fields. We use this methodology to determine the nonlinear CTOD solution and the length scale $\lengthSCd$ that bounds the region around the crack tip in which the nonlinear mechanical response differs significantly from the standard LEFM predictions. Previously, a similar nonlinear length scale defined in terms of a specific ratio of second- to first-order elastic contributions has been proposed~\cite{Bouchbinder2010a, Morishita2016, Mai2023}; in this work, we employ a matched-asymptotics approach~\cite{Geubelle1994} to define $\lengthSCd$ as the physical distance from the crack tip to the crossover between the nonlinear CTOD solution and its classical LEFM prediction. This crossover allows us to express $\lengthSCd$ as a function of the far-field loading and geometric conditions through the stress intensity factor $K_I$.

Building on our observation that GNL alone leads to novel CTOD behavior, we derive analytical solutions of the near-tip elastic fields including the effects of GNL, and we obtain an expression for the nonlinear length scale $\lengthSCd$ as a function of the material parameters and $K_I$. We use $\lengthSCd$ to formally define the CTOD as a piecewise function comprising the asymptotic nonlinear solution and the standard LEFM prediction. Based on the results obtained using the CTOD method and the analytical results, we present insights into the effects of large deformations on crack mechanics and we argue that this GNL framework serves as a canonical basis for the broader study of nonlinear elastic fracture mechanics (NLEFM).

\begin{twocolfig}[!b]
    \centering
\includegraphics[width=1\linewidth]{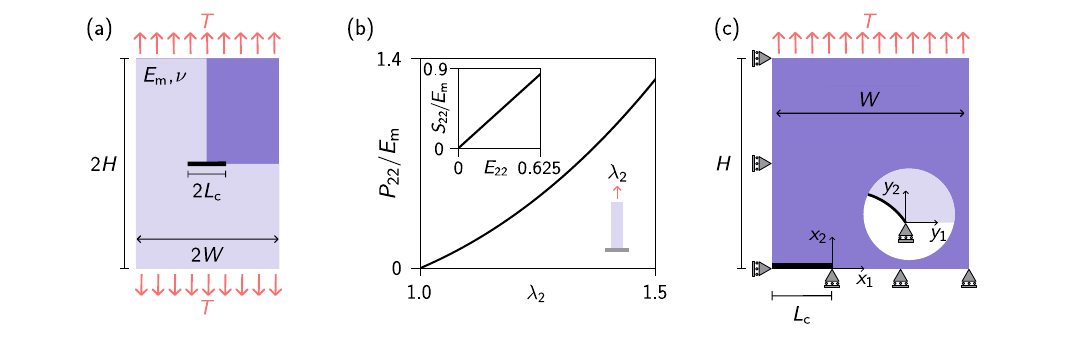}
    \caption{The system considered in this work.
    (a) A plane-strain plate with a central predefined crack under mode I loading. $\E$ and $\nu$ are the Young's modulus and Poisson's ratio of the SVK material, respectively.
    (b) Mechanical response under uniaxial conditions. Nominal stress $P_{22}$ normalized by $\E$ as a function of vertical stretch $\lambda_y$ for $\nu=0.3$. Note the strain-stiffening behavior resulting from nonlinear kinematics. The inset shows the vertical component of the normalized second Piola--Kirchhoff stress $\secondPK[22]/\E$ as a function of Green--Lagrange strain $\greenL[22]$.
    (c) The reduced system considered in the finite element simulations. The origin of the coordinate system for both reference $(x_1,x_2)$ and deformed $(y_1,y_2)$ configurations is the crack tip.}
    \label{fig:setup}
\end{twocolfig}

\section{Problem formulation} \label{sec:problem_formulation}

In this section, we describe the experimental setup, constitutive law, and numerical methods we use to study the effects of GNL on crack behavior in a highly deformable material. 

\subsection{Physical problem}\label{subsect:problem}

Our objective is to provide a fundamental understanding of the effects of GNL on the mechanical response of a static crack undergoing large deformations. To this end, we consider a plate of dimensions $2W\times 2H$ with a predefined central crack of length $2L_\mathrm{c}$ under plane-strain conditions. We assume the plate is made of an SVK material of Young's modulus $\E$ and Poisson's ratio $\nu$. We consider the system under mode I loading with a tensile load $T$ applied at the upper and lower boundaries (see \cref{fig:setup}a). We choose values of $W$ and $H$ to ensure a quasi-infinite plate condition.

We define the strain energy density, $\energyDen$, of a plane-strain compressible SVK material with a stress-free reference configuration, whose domain $\refDomain$ is described with Cartesian coordinates $(x_1, x_2)$ \cite{Holzapfel2000}:
\begin{equation}
\energyDen=\mu \greenL[ij]\greenL[ij]+\frac{\lambda}{2} \greenL[ii]\greenL[jj]\;,
\label{eqn:svk_energy_den}
\end{equation}
where in the above expression we have used Einstein notation, $\lambda$ and $\mu$ are the Lamé coefficients, and $\greenL[ij]$ are the components of the Green--Lagrange strain tensor given by
\begin{equation}
\greenL[ij]=\frac{1}{2}\left(u_{i,j}+ u_{j,i}+u_{k,i}u_{k,j}\right)\;,
\label{eqn:greenL}
\end{equation}
for the displacement vector field $u_i$. Furthermore, we define the second Piola--Kirchhoff stress $\secondPK[ij]$, which is the work conjugate to $\greenL[ij]$, as:
\begin{equation}
\secondPK[ij] = \frac{\pd  \energyDen}{\pd \greenL[ij]} = \lambda \greenL[kk]\id[ij]+2\mu \greenL[ij]\; .
\label{eqn:secondPK}
\end{equation}
We note that in this model the stresses are linearly related to the strains, so that the only source of nonlinearity is the term $u_{k,i}u_{k,j}$ in the strain tensor (\cref{eqn:greenL}), which captures the effects of GNL.

Finally, we define the corresponding boundary value problem (BVP) with respect to the reference configuration. To this end, we first introduce the deformation gradient tensor $\defGrad[ij]$, which maps the material points $(x_1,x_2)$ to spatial points $(y_1,y_2)$ in the deformed domain $\defDomain$, such that $\defGrad[ij]=\pd y_i/\pd x_j$. The first Piola--Kirchhoff stress is then defined via the transformation $\firstPK[ij]=\defGrad[ik]\secondPK[kj]$ \cite{Holzapfel2000}  (see \cref{fig:setup}b). The BVP is then formulated as follows:

\begin{align}
    \firstPK[ij,j] &= 0 \quad \text{in } \refDomain, \label{eqn:BVP0} \\
    \refT[i] &= \firstPK[ij] N_j \quad \text{on } \bound \subset \refDomain, \label{eqn:BVP}
\end{align}
where $T_i$ is a local Piola traction and $N_j$ the normal vector to a boundary $\bound$.

\subsection{Numerical setup}\label{subsect:num_form}

We solve the problem described above numerically using a nonlinear finite-element method implemented in the \verb|FEniCS| computing platform \cite{AlnaesEtal2015}. The BVP defined by \cref{eqn:BVP0,eqn:BVP} is reformulated using a Total Lagrangian approach \cite{Bathe1982}; we provide further details of this implementation in \cref{sect:appendix_TL}. Exploiting the double symmetry of the problem, we model a quarter of the plate (see \cref{fig:setup}c) with a computational domain of dimensions $W\times H$, with $W=H$ and $L_\mathrm{c}=0.05W$; these values ensure an quasi-infinite plate condition.

\section{A CTOD-based numerical analysis}\label{sect:num_analysis}

In this section, we propose a general methodology to characterize the nonlinear response of a crack during the loading process using only its CTOD. Although we use this CTOD-based method specifically to investigate the effects of GNL, it can be applied to a range of problems involving the nonlinear behavior of cracks. Our analysis focuses on the deformed state of the system, which we refer to as the deformed configuration. As shown in \cref{fig:setup}c, we consider the crack tip as the origin of the spatial coordinates $(y_1,y_2)$. We then describe the upper half of the CTOD $\delta$ with the pair $\left(-y_1, \delta\!\left(y_1\right) \right)$, where $-L_\mathrm{c} \leq y_1 \leq 0$ and $\delta\!\left(y_1\right)=y_2$. For the remainder of this study, the term ``CTOD'' will specifically refer to the profile defined by $\left(-y_1, \delta\!\left(y_1\right) \right)$.

We begin by qualitatively comparing the numerically obtained CTOD, including GNL, with Westergaard's analytical solution \cite{Westergaard1939} (see \cref{fig:ctod_gnl_sm}a). We observe that GNL induces a significant stiffening of the plate at high loads, which reduces the extent to which the crack opens under mode I loading. Although these effects are more subtle when the system is subject to lower loads, the influence of GNL can be clearly seen using a log-log scale (see \cref{fig:ctod_gnl_sm}b). This representation is particularly useful because the CTOD follows a power-law relationship with distance from the crack tip, and in this representation the exponent of the power law is given by the slope of the curve. Notably, we see that in the asymptotic region at the crack tip, the slope of the CTOD transitions from the classical LEFM value of $1/2$ (blue lines) to a clearly distinct power-law regime. We note that previous experimental and theoretical studies on PAAm hydrogels have identified similar asymptotic behavior for nonlinear elastic materials \cite{Livne2008,Bouchbinder2008,Goldman2012,GoldmanBoue2015a}.

In the remainder of this section, we first determine the asymptotic characteristics of the CTOD including GNL and identify the region far from the crack tip in which the LEFM still makes accurate predictions. We then introduce and compute the length scale $\lengthSCd$ that defines the transition point between the nonlinear and linear solutions describing the CTOD.

\begin{twocolfig}[!t]
    \centering
\includegraphics[width=1\linewidth]{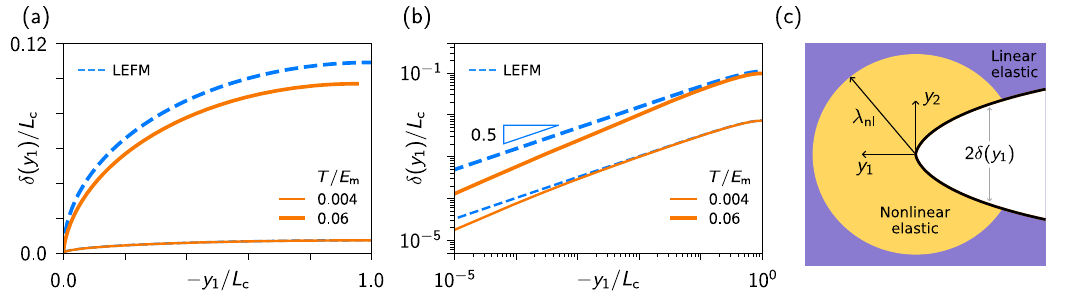}
    \caption{GNL introduces a separation of scales around the crack tip and deviations from LEFM-predicted behavior.
    (a) The normalized CTOD profile for the plate subject to different loads $T$ and $\nu=0.3$. The orange curves correspond to simulations including GNL, and blue dashed lines correspond to Westergaard's analytical solution (LEFM).
    (b) The normalized CTOD profile on a log-log scale showing the same data as in (a).
    (c) Schematic of the scale separation in the system. Within a distance $\lengthSCd$ of the crack tip, nonlinear elasticity controls the mechanical response of the system. Outside this region, deformations are small and the response can be assumed to be linear. We note that the circular shape of the nonlinear region is only illustrative; in practice, the nonlinear length scale may exhibit a dependence on angular direction.
}
    \label{fig:ctod_gnl_sm}
\end{twocolfig}

\subsection{Asymptotic nonlinear behavior}\label{subsect:asymptotic_behavior}

As shown in \cref{fig:ctod_gnl_sm}b, close to the crack tip a power-law relationship distinct from that predicted by LEFM can be observed. The region in which this power law dominates can be seen to grow as the loading on the system is increased. To discern this asymptotic exponent accurately from our numerical experiments, we consider a range of $T/\E$ values and analyze the scaling of $\delta$ with $y_1$. As shown in \cref{fig:slopes}a (see also \cref{fig:appendix_slopes}), when approaching the crack tip, the CTOD scaling transitions from $\delta \propto (-y_1)^{1/2}$ to $\delta \propto (-y_1)^q$; where this transition occurs is dependent on the loading. We note that the nonlinear behavior is always present sufficiently close to the crack tip, and we observe a unique value of $q$ for a given material. However, considering large loads on the system permits us to evaluate $q$ while avoiding inaccuracies associated with a finite mesh resolution. Furthermore, we note that the asymptotic exponent is not universal but instead depends on $\nu$, as shown in \cref{fig:slopes}b. Indeed, as $\nu$ approaches the incompressibility limit, the material constrains both the crack-tip opening and the horizontal deformation more severely than when $\nu$ is close to zero, leading to an exponent $q$ that is closer to the LEFM value of $1/2$. Details of the methodology used here are presented in \cref{sect:appendix_ctod_1}.

\begin{twocolfig}[!t]
    \centering
\includegraphics[width=1\linewidth]{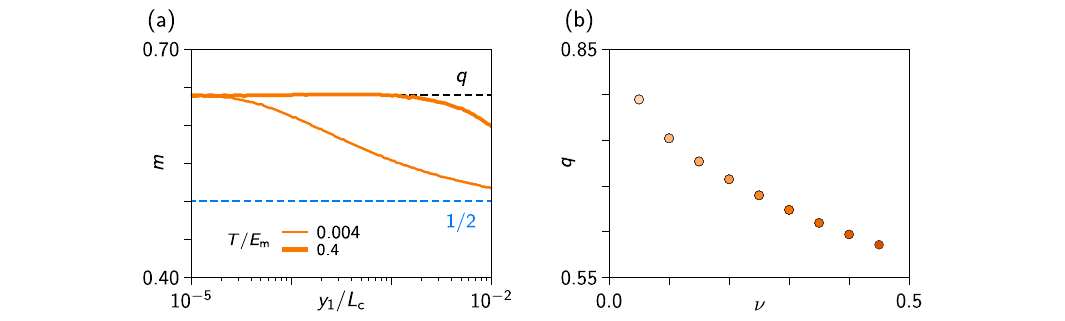}
    \caption{
    Computation of nonlinear asymptotic exponent. 
    (a) Local exponent $m(y_1)$ of the CTOD defined by $m(y_1)=d\log(y_2)/d\log(y_1)$ (see \cref{fig:ctod_gnl_sm}b) as a function of the horizontal distance from the crack tip for different loads $T$ ($\nu=0.3$). The blue dashed line corresponds to a slope of $1/2$ and the black dashed line is the final value indicating convergence to the asymptotic exponent $q$.
    (b) The asymptotic exponent $q$ as a function of $\nu$.
}
    \label{fig:slopes}
\end{twocolfig}

\subsection{Nonlinear length scale}\label{subsect:nl_lengthscale}

Having identified two regions in which two distinct power laws describe the CTOD, we now compute the nonlinear length scale that defines the transition between these two domains. We refer to the regions in which the nonlinear and linear responses dominate as the inner and outer regions, respectively.

The nonlinear length scale $\lengthSCd$ defines the region around the crack tip in which the CTOD is dominated by the asymptotic exponent $q$ and demarcates the transition from the inner to the outer region (see \cref{fig:lambda_nl_cur}a). Here, we define $\lengthSCd$ as the horizontal coordinate of the crossover point where the inner and outer power-law solutions intersect (i.e., $\delta^\text{inner}(\lengthSCd)=\delta^\text{outer}(\lengthSCd)$, see \cref{fig:lambda_nl_cur}a). Using the appropriate value of $q$ (\cref{fig:slopes}b), we fit the CTOD data to both power-law functions and find their intersection to determine the nonlinear length scale. We note that due to the piecewise nature of the CTOD considered here, and the fact that the outer solution is given by LEFM, we are able to relate the behavior of the whole CTOD, including the nonlinear asymptotic region, to the far-field loading and geometric conditions. Because we assume the validity of LEFM in the outer region, we directly compute the stress intensity factor from the CTOD and denote it as $K_I^{(\delta)}$. We provide further details in \cref{sect:appendix_ctod_2}.

In \cref{fig:lambda_nl_cur}b, we plot $\lengthSCd/L_\mathrm{c}$ as a function of $K_I^{(\delta)}/\mu$. For small loading, we observe the relationship $\lengthSCd\sim \left(K_I^{(\delta)}/\mu\right)^2$ (see \cref{fig:lambda_nl_cur}b, inset). Furthermore, we find that the dependence of $\lengthSCd$ on $\nu$ is non-monotonic; for a given value of $K_I^{(\delta)}/\mu$, $\lengthSCd$ exhibits a maximum value in the range $0.2 < \nu < 0.3$, and $\lengthSCd$ decreases as $\nu \to 0$ or $\nu \to 0.5$. To understand the origin of this behavior, we first note that the LEFM solutions assume that the horizontal deformation of the crack is zero. Considering the opening of the crack while suppressing horizontal deformations, as detailed in \cref{sect:appendix_sref} and referred to as the pseudo-reference configuration, we observe that the asymptotic behavior of the vertical and horizontal deformations scale differently with respect to the reference position. It can be shown that the Poisson-mediated coupling between these distinct directional scaling laws drives the non-monotonic behavior.

\begin{twocolfig}[!t]
    \centering
\includegraphics[width=1.0\linewidth]{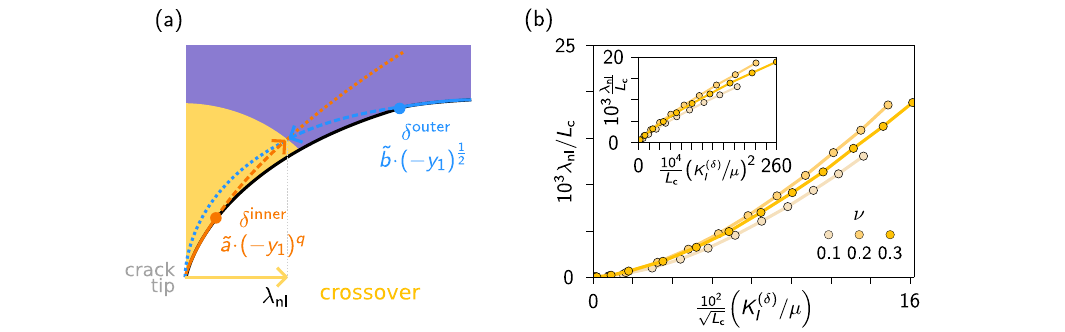}
    \caption{Computation of nonlinear length scale $\lengthSCd$.
    (a) Schematic of the proposed solution. The inner solution given by $\delta^\text{inner}=\tilde{a} \cdot (-y_1)^{q(\nu)}$ and the outer solution is $\delta^\text{outer}=\tilde{b} \cdot (-y_1)^{1/2}$, where $\tilde{a}$ and $\tilde{b}$ are unknown positive coefficients. Note that the origin of both power laws is defined as the crack tip.
    (b) Numerical results of $\lengthSCd/L_\mathrm{c}$ as a function of $K_I^{(\delta)}/\mu$. The inset shows the scaling of $\lengthSCd/L_\mathrm{c}$ with respect to $(K_I^{(\delta)}/\mu)^2$.
}
    \label{fig:lambda_nl_cur}
\end{twocolfig}

\section{Analytical asymptotic solution}
\label{sect:asymptotic_fields}

Motivated by our numerical findings, in this section we present an asymptotic analysis within a fully nonlinear elastic framework leading to an analytical solution of the static crack-tip fields including the effect of GNL. 

\subsection{Geometrically nonlinear crack problem}\label{subsect:gov_eqs}

We start by restating the problem we analyze analytically. We consider a plane-strain isotropic GNL solid; the reference configuration is stress free, with points occupying the two-dimensional domain $\refDomain$ described by the Cartesian coordinates $x_i=\left(x_1,x_2\right)$. Its deformed state, described by $y_i=\left(y_1,y_2\right)\in\defDomain$, is obtained via the displacement field $u_i$, i.e., $y_i=x_i+u_i$. Then, the deformation gradient that maps points from the reference configuration to the deformed configuration is defined as $\defGrad[ij]=\pd y_i/\pd x_j$, and its determinant, describing volumetric changes, corresponds to $\jacDet=\Det \defGrad[ij]$.

We include GNL within the model by considering the plate to be an SVK material, as introduced in \cref{eqn:svk_energy_den}. To aid the derivation of an explicit expression for the nominal stresses $\firstPK[ij]$, we use the following plane-strain form of the SVK model:
\begin{equation}
\energyDen(\firstI,\jacDet)=\frac{\mu}{2} \left( \firstI-\jacDet^2-1\right)+\kappa\frac{\mu}{4}\left(\firstI-2\right)^2 \;,
\label{eqn:svk_energy_den_IJ}
\end{equation}
with the parameter $\kappa=\frac{\lambda+2\mu}{2\mu}=\frac{1-\nu}{1-2\nu}$, and where $\rightCG[ij]$ is the right Cauchy--Green strain, $\rightCG[ij]=\id[ij]+2\greenL[ij]$, and $\firstI$ its first invariant, $\firstI=\rightCG[kk]$. The equivalence between \cref{eqn:svk_energy_den,{eqn:svk_energy_den_IJ}} follows from the identity between the invariants under plane-strain assumption $2\jacDet^2=\firstI^2-\rightCG[ij]\rightCG[ij]$. We utilize \cref{eqn:svk_energy_den_IJ} to obtain the first Piola–Kirchhoff stress $\firstPK[ij]$ as:
\begin{equation}
\firstPK[ij] = \frac{\pd \energyDen}{\pd \defGrad[ij]}= \mu\left((1-2\kappa+\kappa \firstI)\defGrad[ij] - \jacDet\epsilon_{i k}\epsilon_{j l} \defGrad[kl]\right)
\;,
\label{eqn:pk1_alternative}
\end{equation}
where $\epsilon_{ij}$ is the two-dimensional permutation symbol \cite{Holzapfel2000}. Finally, we use the nominal stresses defined in \cref{eqn:pk1_alternative} in the BVP problem already presented in \cref{eqn:BVP0,eqn:BVP}.

We consider $\refDomain$ as the domain occupied by an undeformed infinite plate containing a semi-infinite crack defined by the region $ x_1 < 0$ and $x_2=0^\pm$. Since the crack faces must be traction-free, the pseudo-stresses $P_{ij}$ satisfying the equilibrium condition \cref{eqn:BVP} are subject to homogeneous boundary conditions:
\begin{equation}
\firstPK[i2] (x_1<0,x_2=0^\pm)=0 \;.
\label{eq:bc1}
\end{equation}
Furthermore, we assume that the body is loaded at its boundaries such that mode~I loading is ensured.

\subsection{The inner solution}\label{subsect:inner_sol}

Our numerical experiments show the existence of an inner region where GNLs are dominant and dictate the mechanical response of the crack; we now aim to define an analytical form of the CTOD in this inner region. Let us consider a polar coordinate system defined with its origin at the crack tip in the reference configuration, i.e., $x_1=r\cos\theta$ and $x_2=r\sin\theta$, with $\theta \in \left[-\pi,\pi\right]$. We assume the deformation of the region close to the crack tip takes the following asymptotic form \cite{Knowles1973}:
\begin{equation}
y_{i}(r,\theta) \simeq a_i^{1-m_i}\angSoll[i](\theta)\, r^{m_{i}}\;,
\label{eqn:sol_form}
\end{equation}
with $\angSoll[i]\left(\theta\right)$ being angular functions and $m_i$ and $a_i$ are positive scalar values; we note that $a_i$ have dimensions of length. Furthermore, as shown in Knowles and Sternberg \cite{Knowles1973, Knowles1974}, the exponents $m_i$ must satisfy the following conditions to be physically relevant:
\begin{equation}
m_1 > m_2 \hspace{10pt}\text{and}\hspace{10pt} m_2 <1 \;.
\label{eqn:conds}
\end{equation}
To fully define the inner solution, we must obtain the constants $m_i$ and $a_i$ as well as the functional form of $\angSoll[i]\left(\theta\right)$. 

We begin by obtaining an expression for the exponent $m_2$. To this end,
we introduce the J-integral in polar coordinates as follows:
\begin{equation}
\Jint = \int_{\closeC}  \left(\energyDen n_{1}-\firstPK[ij ] n_{j} u_{i,1}\right) r \nd \theta \;,
\label{eqn:Jint}
\end{equation}
where $\closeC$ is a closed contour in $\refDomain$ enclosing the crack tip. We examine the singular behavior of the integrand of \cref{eqn:Jint} as $r\to 0$ to identify that the dominant term diverges as $r^{4(m_2-1)}$ 
. We then use the integrability condition of $\Jint$ to find that to avoid it vanishing as $r\to 0$, the condition $4(m_2-1)=-1$ must be satisfied. Therefore, we obtain the value of the exponent $m_2$ as: 
\begin{equation}
m_2=\frac{3}{4}\;.    
\label{eqn:m_2}
\end{equation}

\begin{figure}[!t]
\begin{center}
\includegraphics[width=1.0\linewidth]{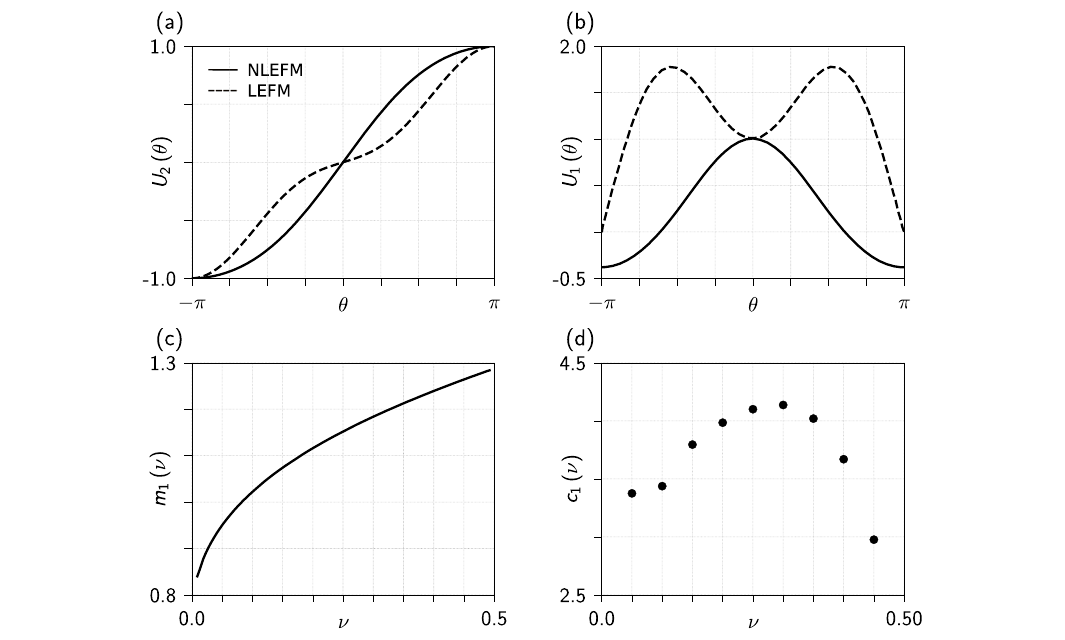}
\caption{Angular functions and parameters for semi-analytical solution. The solid lines represent the nonlinear elastic solution and the dashed lines represent the LEFM solutions.
(a) Analytical $U_2(\theta)$ with $\nu=1/3$ and $m_2=3/4$.
(b) Numerically obtained values of $U_1(\theta)$ for $\nu=1/3$ and $m_1 = 1.20279$.
(c) Eigenvalues, $m_1(\nu)$, of \cref{eqn:eigen_problem} computed numerically.
(d) Values of $c_1(\nu)$, defined in \cref{eqn:def_c1}, and obtained by fitting of $y_1$ to FE results.
}
\label{fig:theory}
\end{center}
\end{figure}

We now aim to find the solution for $\angSoll[2]\left(\theta\right)$. Considering the equilibrium equation $\firstPK[2 j,j]=0$ (\cref{eqn:BVP}), we use the diverging behavior $\firstPK[2j]\propto \firstI \defGrad[2j]$ to find its asymptotic form (see derivation in \cref{subsect:appendix_eqU2}); we find that
\begin{equation}
\left(\ddot{U}_2+m_2^2\angSoll[2]\right)\angOne+2m_2\left(m_2-1\right)\angSoll[2]\,\angOne+\dot{U}_2\dot{\angOne} =0
\;,
\label{eqn:eq_U2}
\end{equation}
where the dot denotes differentiation with respect to $\theta$ and $\angOne(\theta) = m_2^2U_2^2+\dot{U}_2^2$.
We then obtain the following boundary conditions that can be used in \cref{eqn:eq_U2}: $\angSoll[2]\!\left(0\right)=0$, $\dot{U}_2\!\left(\pm \pi\right)=0$, and $\angSoll[2]\!\left(\pm\pi\right)=\pm 1$. The first condition comes from the antisymmetry of $y_2$ under mode I loading, i.e., $U_2\!\left(\theta\right) = - U_2\!\left(-\theta\right)$. The second is a result of the vertical stress-free state condition along the crack (\cref{eq:bc1}), and the third boundary condition is a normalization condition for $\angSoll[2]\!\left(\theta\right)$.
The solution of \cref{eqn:eq_U2} with these boundary conditions has been discussed previously in the literature in a similar context \cite{Knowles1973, Geubelle1994, Long2011}, and it is given by:
\begin{equation}
U_2(\theta)=\frac{{\rm sgn}(\theta)}{2}\left(\omega(\theta)-2\cos\theta\right)^{1/2}\,\left(\omega(\theta)+\cos\theta\right)^{1/4}\;,
\label{eqn:U2}
\end{equation}
where ${\rm sgn}(\theta)=\theta/|\theta|$ and
\begin{equation}
\omega(\theta)=\sqrt{3+\cos^2\theta} \;   \;.
\end{equation}
In \cref{fig:theory}a the angular variation $\theta=\left[-\pi,\pi\right]$ of the normalized quantity $U_2$ is shown and compared with the equivalent LEFM result.

The conditions arising from the asymptotic analysis are not sufficient to determine the coefficient $a_2$. To obtain an expression for this parameter, we make use of the far-field loading and geometric conditions. We utilize the path-independence of $\Jint$: defining two arbitrary contours, one in the inner region and a second in the outer region, we obtain  $\Jint_\text{inner}=\Jint_\text{outer}$. Considering that $\Jint$ is a measure of the energy release rate $G$ of the system, the last equality is equivalent to $G_\text{inner} = G_\text{outer}$. We thus use \cref{eqn:Jint} to compute $G_\text{inner}$:
\begin{equation}
    G_\text{inner}= \frac{27\pi}{128}  \kappa\mu\, a_2 \;.
    \label{eqn:Ginner}
\end{equation}
In the outer region, we assume that the LEFM solution is valid. Then, we use Irwin's relationship to compute $G_\text{outer}$, which for plane-strain conditions corresponds to \cite{Broberg1999}:
\begin{equation}
    G_\text{outer}=\frac{\left(1-\nu\right)\, K_I^2}{2\mu} \;.
    \label{eqn:Gouter}
\end{equation}
\cref{eqn:Ginner,eqn:Gouter} can then be used to find $a_2$:
\begin{equation}
    a_2 = \frac{64}{27\pi }\frac{1-\nu}{\kappa}\, \left(\frac{K_I}{\mu}\right)^2
    \;,
    \label{eqn:a_2}
\end{equation}
which completes the derivation of the analytical solution for $y_2=a_2^{1-m_2}U_2(\theta)\, r^{m_2}$. This solution for $y_2$ provides an analytic solution for the CTOD with the pair $(r,y_2)$, which corresponds to a pseudo-reference configuration; further analysis of the system in this reference frame is presented in \cref{sect:appendix_sref}.

In the case of the solution $y_1$, a solution for $m_1$ and $U_1$ should be determined numerically. To this end, we use the equilibrium equation $\firstPK[1 j,j]=0$ and derive its asymptotic form as (see details in \cref{subsect:appendix_eqU1}):
\begin{equation}
\kappa \left( \left(m_1+2m_2-2\right)m_1U_1\angOne+\dot{U}_1\,\dot{\angOne}+\ddot{U}_1\,\angOne\right)=
 (m_1+m_2-2)\dot{U}_2\,\angTwo-m_2U_2 \,\dot{\angTwo}\;\; ,
\label{eqn:eigen_problem}
\end{equation}
where $\angTwo(\theta) = m_1 \angSoll[1] \dot{U}_2-m_2 U_2 \dot{U}_1$. We highlight that in \cref{eqn:eigen_problem} neither of the coefficients $a_1$ and $a_2$ appear. Furthermore, we impose the boundary conditions $\dot{U}_1\!\left(0\right)=0$, $\dot{U}_1\!\left(\pm\pi\right)=0$, and $U_1\!\left(0\right)=1$. These conditions arise from the symmetry of $y_1$, i.e.,  $U_1\!\left(\theta\right)=U_1\!\left(-\theta\right)$; the stress-free condition $P_{12}\!\left(\pm\pi\right)=0$; and the normalization condition for $\angSoll[1]$, respectively. Finally, we numerically solve \cref{eqn:eigen_problem} considering these boundary conditions as an eigen-problem for $m_1$ and $U_1$, where the solution is conditioned by the chosen value of $\nu$, such that $m_1=m_1\!\left(\nu\right)$. In \cref{fig:theory}b-c, we plot $U_1$ and $m_1$, respectively, and compare the first against the equivalent LEFM results.

Finally, we aim to determine the form of the missing coefficient $a_1$ using scaling arguments. As shown in \cref{eqn:a_2}, $a_2$ scales with $(K_I/\mu)^2$. Since the SVK material law introduces no intrinsic length scale, the quantity $(K_I/\mu)^2$ represents the only available length scale for the inner asymptotic problem. Given that both $a_i$ coefficients have dimensions of length (see \cref{eqn:sol_form}), dimensional consistency requires that
\begin{equation}
  a_1 = c_1(\nu)^{\frac{1}{1-m_1}} \, \left(\frac{K_I}{\mu}\right)^2\;,  
  \label{eqn:def_c1}
\end{equation}
where $c_1(\nu)$ is a prefactor to be determined. In principle, the functional form of $c_1(\nu)$ could be determined by considering higher-order terms in the asymptotic expansion of $y_i$ and ensuring that the coefficients $a_1$ and $a_2$ do not vanish in the equilibrium equation $P_{1j,j} = 0$ \cite{Knowles1974, Geubelle1994}. In the present work, we do not undertake this higher order asymptotic analysis, but instead combine theoretical and numerical results. We obtain values for $c_1(\nu)$ by fitting $y_1(r,\pi)$ to CTOD data (see \cref{fig:theory}d), which together with the solutions to \cref{eqn:eigen_problem} fully define $y_1$.

\subsection{Nonlinear crack-tip opening displacement}\label{subsect:matched_asymp}

We have shown in \cref{sect:num_analysis} that the CTOD consists of two distinct power-law solutions and that the transition between the two solutions occurs at a distance $\lengthSCd$ from the crack tip. Using the results of the previous section, here we formally define the CTOD as a piecewise function of the inner and outer solutions and obtain an expression for $\lengthSCd$. Given that we have obtained a complete analytical solution for $y_2=y_2(r,\theta)$, we first define the nonlinear length scale in the pseudo-reference configuration (i.e., neglecting horizontal deformations). In the pseudo-reference frame, we denote the nonlinear length scale and the CTOD profile as $\lengthSCr$ and $\ctodr$, respectively. A description of the system in this reference frame is presented in \cref{sect:appendix_sref}.

In the vicinity of the crack tip, the CTOD in the pseudo-reference configuration is obtained when $\theta=\pm\pi$ so that $r=-x_1$. Noting that $|U_2(\pm\pi)|=1$ and using \cref{eqn:m_2}, we see that the inner solution for the CTOD corresponds to $\innersolr=a_2^{1/4}\, r^{3/4}$, where $a_2$ is given by \cref{eqn:a_2}. We define the outer solution from LEFM as $\outersolr = b^{1/2}\, r^{1/2}$, where for plane-strain conditions \cite{Anderson2017}:
\begin{equation*}
b=\frac{2}{\pi}\left(1-\nu\right)^2\, \left(\frac{K_I}{\mu}\right)^2
\;.
\end{equation*}
By applying $\theta = \pm\pi$ so that $|U_2|=1$ and  $r = -x_1$, we define the CTOD in the pseudo-reference configuration as:
\begin{equation}
    \ctodr \left(-x_1\right) = 
                         \left\{
                         \begin{array}{ll}
                          a_2^{1/4}\, \left(-x_1\right)^{3/4} &  \mbox{for}\quad 0 < -x_1 \ll \lengthSCr
                          \\
                          b^{1/2}\, \left(-x_1\right)^{1/2} &  \mbox{for}\quad -x_1 \gg \lengthSCr
                        \end{array}
                        \right.
\;,
\label{eqn:CTOD_psref}
\end{equation}
where $\lengthSCr$ is the critical transition value. We note that based on the definition of the CTOD used in this work, \cref{eqn:CTOD_psref} corresponds to half of the total opening of the crack (see \cref{fig:ctod_gnl_sm}c). Furthermore, we obtain an expression for $\lengthSCr$ by evaluating the crossover between the inner and outer solutions at $-x_1 \to  \lengthSCr$, such that $\innersolr (\lengthSCr)=\outersolr(\lengthSCr)$. We use the previous results to solve this crossover problem finding the analytical expression for the nonlinear length scale:
\begin{equation}
    \lengthSCr = \frac{27}{16\pi}\frac{\left(1-\nu\right)^4}{1-2\nu}\left(\frac{K_I}{\mu}\right)^2    
    \;.
    \label{eqn:lambda_nl_r}
\end{equation}

Furthermore, we proceed by defining the CTOD and nonlinear length scale in the deformed configuration. We begin by using \cref{eqn:sol_form} combining the components $y_1$ and $y_2$ to define the inner solution of the CTOD as $\innersold (-y_1) = a^{1-q}\cdot\left(-y_1\right)^q$, where $a$ and $q$ are scalar coefficients defined below. Since the LEFM framework is insensitive to the material configuration, we use $\outersold = \outersolr$. We thus define the CTOD in the deformed configuration:
\begin{equation}
    \ctodd\!\left(-y_1\right) = 
                         \left\{
                         \begin{array}{ll}
                          a^{1-q}\, \left(-y_1\right)^{q} &  \mbox{for}\quad 0 < -y_1 \ll \lengthSCd
                          \\
                          b^{1/2}\, \left(-y_1\right)^{1/2} &  \mbox{for}\quad -y_1 \gg \lengthSCd
                        \end{array}
                        \right.
                        \;,
\label{eqn:ctod_def}                   
\end{equation}
which corresponds to half of the total crack opening $2\ctodd$. To define the exponent $q$, we use the analytical result $m_2$ (\cref{eqn:m_2}) and the eigenvalues $m_1$ (\cref{fig:theory}b), finding that:
\begin{equation}
q \equiv \frac{m_2}{m_1}= \frac{3}{4m_1}\;.    
\end{equation}
We then define the coefficient $a$ using the analytical expression for $a_2$ (\cref{eqn:a_2}) and recalling that $a_1$ depends on the prefactor $c_1(\nu)$ to obtain:
\begin{equation}
a = \frac{\left(\frac{64}{27\pi}(1-2\nu)\right)^{\frac{1}{4(1-q)}}}{ \left(c_1(\nu)\left\vert{}U_1(\pi)\right\vert{}\right)^{\frac{q}{1-q}} }\, \left(\frac{K_I}{\mu}\right)^2
\;.
\label{eqn:a}
\end{equation}
We now derive an expression for the nonlinear length scale $\lengthSCd$. To this end, we use the previous results to identify the crossover in the deformed configuration, $\innersold = \outersold$, and solve for $\lengthSCd$. We thus define the nonlinear length scale as:
\begin{equation}
    \lengthSCd = h\!\left(\nu\right)\cdot \left(\frac{K_I}{\mu}\right)^2
    \;,
    \label{eqn:length_sc_d}
\end{equation}
with the prefactor:
\begin{equation}
    h(\nu) = \left( \frac{27}{16\pi} \frac{(1-\nu)^4}{(1-2\nu)} \right)^{\frac{1}{4q-2}} \left(c_1(\nu)\left\vert{} U_1(\pi) \right\vert{}\right)^{\frac{2q}{2q-1}}    \;.
    \label{eqn:h}
\end{equation}
We remark that while $m_1$ (and thus $q$ in \cref{eqn:ctod_def}) and $U_1$ in \cref{eqn:sol_form} are obtained from \cref{eqn:eigen_problem}, to fully specify $\lengthSCd$ and the CTOD solution, the prefactor $c_1$ must be obtained from the fitting of $y_1$ to the CTOD data in the pseudo-reference configuration.

\section{Theoretical and numerical validation}
\label{subsect:validation}

Here, we validate our results by comparing the theoretical solutions presented in  \cref{sect:asymptotic_fields} against the numerical results obtained in \cref{sect:num_analysis}. Specifically, we evaluate the agreement between the two methodologies in terms of the asymptotic exponent $q$ of the CTOD and the nonlinear length scale $\lengthSCd$, both depicted schematically in \cref{fig:lambda_nl_cur}a. 

\begin{figure}[!t]
    \centering
\includegraphics[width=1\linewidth]{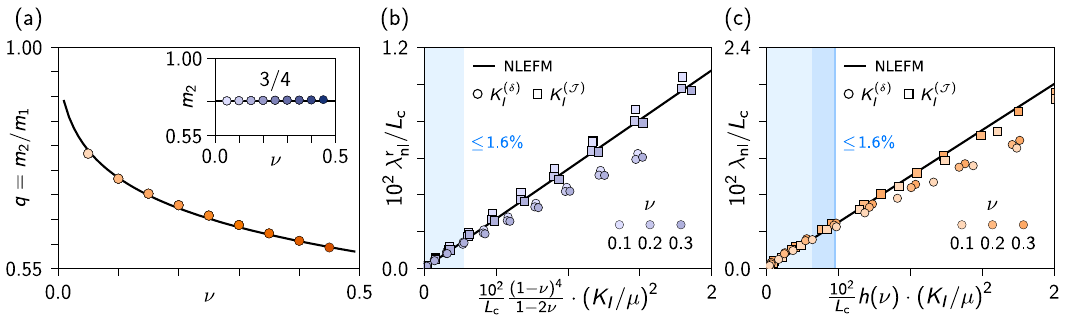}
    \caption{
Comparison between theoretical predictions for the CTOD and numerical results. Solid black lines denotes the NLEFM theory presented in this work (\cref{sect:asymptotic_fields}), while points represent numerical FE results (\cref{sect:num_analysis}).
(a) The values of the asymptotic exponent $q$ as a function of Poisson's ratio, $\nu$, in the deformed configuration. Inset: Results for the pseudo-reference configuration ($m_2 = 3/4$).
(b and c) Normalized nonlinear length scale versus the square of the normalized stress intensity factor in the pseudo-reference and deformed configurations, respectively.
Circles represents results obtained using $K_I$ values obtained from the CTOD fitting, $K_I^{(\delta)}$, while squares denote values obtained from the $J$-integral via Irwin’s relationship (\cref{eqn:Gouter}), $K_I^{(\mathcal{J})}$.
The blue shaded regions in both (b) and (c) show the region $T/\E \leq 0.016$, where fully CTOD-based results show good agreement with theoretical predictions; in (c), the size of the region $T/\E \leq 0.016$ varies for each $\nu$.
}
    \label{fig:validation}
\end{figure}

As shown in \cref{fig:validation}a, the numerical and theoretical methods show excellent agreement in terms of the scaling of the CTOD with distance from the crack tip.
To assess the extent to which the methods agree in terms of the nonlinear length scale $\lengthSCr$, we first compare the analytically obtained values of $\lengthSCr$ (\cref{eqn:lambda_nl_r}) with those obtained numerically considering the pseudo-reference configuration (see \cref{sect:appendix_sref}), plotting $\lengthSCr$ against $ (K_I/\mu)^2$ for different values of $\nu$ (see \cref{fig:validation}b). We observe that when using the expression for $K_I$ obtained using the CTOD method ($K_I^{(\delta)}$, see \cref{sect:appendix_ctod_2}), the results collapse onto a master curve, but agreement between the numerical and theoretical results is limited to the region $T/\E < 0.016$. For high loading values, our analytic results deviate from the results obtained via FE simulations. We attribute these inaccuracies to the assumption that far from the crack tip, the material behavior is exactly linear; for large values of loading, this is known not to be true. When using the $\Jint$-integral to obtain $K_I$ (\ie, using \cref{eqn:Jint,eqn:Gouter} to compute $K_I^{(\Jint)}$), we obtain excellent agreement between the theoretical prediction and the numerical results for higher values of $T/\E$ (see \cref{fig:validation}b).

Finally, we assess the accuracy of the predictions of the nonlinear length scale in the deformed configuration. We note that the numerical work as well as the inner solution presented in \cref{subsect:matched_asymp} includes both horizontal and vertical deformations, while the outer solution (\cref{subsect:matched_asymp}), representing the predictions of LEFM,
does not include horizontal displacement along the crack surface.
In \cref{fig:validation}c, we plot the value of $\lengthSCd$ found numerically as a function of $(K_I/\mu)^2$ against the semi-analytical expressions derived in \cref{subsect:matched_asymp}, where the prefactor $h(\nu)$ is given by \cref{eqn:h} (the fitting parameter $c_1(\nu)$ being extracted from FE simulations, see \cref{fig:theory}d). As expected, our theoretical predictions give a master curve, and the accuracy of the results when using $K_I^{(\delta)}$ or $K_I^{(\Jint)}$ are consistent with the previous discussion.

\section{Discussion}
\label{sect:discussion}

In this section, we discuss the implications of our theoretical and numerical findings, the limitations and sources of inaccuracy in the work presented here, and the possible generalizations of the current framework to the broader study of NLEFM.

We start by considering why our predictions for $\lengthSCr$ and $\lengthSCd$ are less accurate for large values of $T/\E$ when using $K_I^{(\delta)}$, an observation that motivates us to consider $K_I^{(\mathcal{J})}$. The quantity $K_I^{(\delta)}$ is obtained theoretically by considering the outer solution, where it is assumed that LEFM accurately predicts the CTOD, \ie, we assume that the system is perfectly linear in the region $-y_1 \gg \lengthSCr$. In the case of the numerical experiments, the SVK material law is defined over the entire domain; therefore, although GNL effects are weak in the outer region, the nonlinear terms in the strain definition (\cref{eqn:greenL}) are not identically zero. Neglecting the small nonlinearities present in the outer region thus contributes to the inaccuracies in the estimations of $K_I^{(\delta)}$. Furthermore, for larger values of $T/\E$, the nonlinear length scale increases and the outer region is progressively confined to a region that approaches the crack tail, where it may be influenced by the symmetry boundary at $x_1 = -L_c$. The effect of this boundary, which is not taken into account theoretically, is hypothesized to contribute to the error in the estimate of $K_I^{(\delta)}$. One potential solution for increasing the accuracy of the $K_I^{(\delta)}$ predictions from the CTOD method is to include higher-order terms in the fitting of the outer solution; this fitting, however, in the absence of additional constraints, lacks physical significance. It is for this reason that we consider an alternative derivation of $K_I$ via the $\Jint$-integral, a global measure of the energy release rate. Our results in \cref{fig:validation}b suggest that the $\Jint$-integral accurately captures the behavior of $K_I$ in the outer region over a wider range of applied $T/\E$ values. We note that the $\Jint$-integral approach involves a heuristic extension of Irwin’s relationship incorporating the weak GNL effects captured by $\Jint$, and that it will eventually fail as bulk deformations increase.

\begin{figure}[!t]
    \centering
\includegraphics[width=1\linewidth]{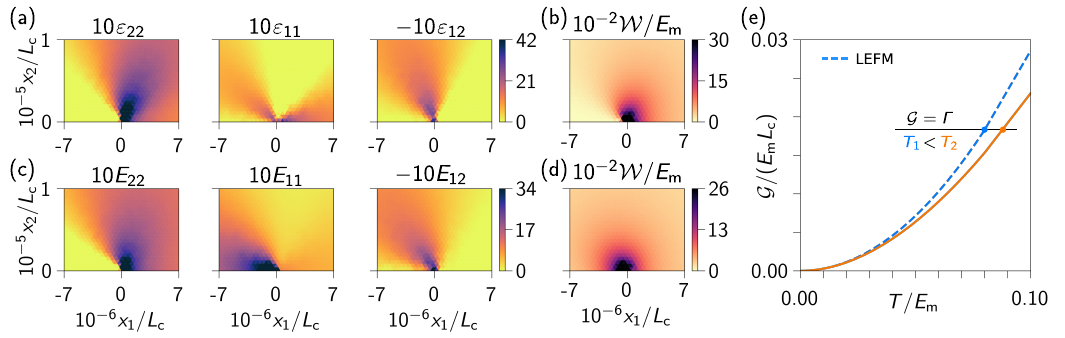}
    \caption{
Evaluation of strain fields, energy density, and energy release rate for $\nu = 0.3$ both with and without GNL. (a and c) The infinitesimal strain components ($\varepsilon_{22}, \varepsilon_{11}, \varepsilon_{12}$), corresponding to the elastic behavior under small deformations, and the Green--Lagrange strain components ($E_{22}, E_{11}, E_{12}$), which include GNL effects. (b and d) The normalized strain energy density distributions for linear elastic and SVK material models, respectively. All contour plots depict the fields within the same region in the reference configuration for $T/\E = 0.004$. (e) The normalized energy release rate $\err$ as a function of the applied load $T/\E$. The dashed blue line indicates the LEFM prediction, the orange curve corresponds to the NLEFM model, and the horizontal black line denotes a theoretical fracture threshold where $\mathcal{G}$ is equal to the fracture energy $\fractEn$. We used the $\Jint$-integral as the measure of $\err$. The infinitesimal strain field, representing LEFM prediction, was obtained from FE simulations following the same setup defined in \cref{fig:setup}c.
}
    \label{fig:strain_and_energies}
\end{figure}

In this work, we have demonstrated that GNL has a significant effect on the crack-tip behavior even for moderate loading. Indeed, as shown in  Figs. \ref{fig:strain_and_energies}a and c, even under a moderate load of $T/\E = 0.004$, the strain fields observed when considering the effects of GNL differ significantly from LEFM predictions. We find that the characteristic V-shape of the horizontal infinitesimal strain $\varepsilon_{11}$ observed in the LEFM results is not present when the effects of GNL are considered, and that it is replaced by a highly localized positive strain $\greenL[11]$ (see \cref{fig:strain_and_energies}c and the angular distribution in \cref{fig:theory}c); furthermore, both $\greenL[22]$ and $\greenL[12]$ exhibit larger values in localized regions behind the crack tip than $\varepsilon_{22}$ and $\varepsilon_{12}$ (\cref{fig:strain_and_energies} and Figs.~\ref{fig:theory}a and c). This results in higher levels of stored elastic energy behind the crack tip (see Figs.~\ref{fig:strain_and_energies}b and d). We correlate these observations by noting that the nonlinear length scale $\lengthSCd$ becomes nonzero as soon as the material is loaded (see \cref{fig:validation}b-c); this implies that physical changes at the crack tip due to large-deformation effects start under what is commonly accepted as the small-deformation regime.

For larger values of applied load, we see that when GNL effects are considered the material exhibits a strain-stiffening response (see \cref{fig:setup}b). This is an intrinsic material response driven by the nonlinear kinematics of finite strains, and it is fundamentally distinct from the stiffening response of constitutively nonlinear elastic solids \cite{Jaspers2014,Raayai-Ardakani2019,Kaur2026}. We show in \cref{fig:strain_and_energies}e that, due to this GNL-induced stiffening, the energy release rate of the system, $\err$, increases more slowly with applied quasi-static loading than in the LEFM case. This results in an apparent toughening of the material, \ie, the material resists significantly higher loads before reaching its critical fracture energy $\fractEn$, and consequently, the system has a larger energetic reservoir at the moment of failure. 
The consequences of this energetic state were studied in previous work \cite{Pundir2024}, where using a SVK material it was shown that, driven by GNL, a crack can propagate at supershear speeds, while the size of the cohesive zone remains finite. We believe that understanding how the nonlinear length scale $\lengthSCd$ established in this work evolves during fracture and its relationship with the finite-sized cohesive zone may lead to significant insight into fracture phenomena in nonlinear systems.

While we have presented the application of the CTOD-based methodology to a centered crack under mode I loading condition, since analytical formulas for $K_I$ are also known for various system configurations, the method is also applicable to other loading cases.
Furthermore, given an analytical expression for $\lengthSCd$ (as derived in \cref{subsect:matched_asymp}) and the fact that the CTOD can be measured directly in experiments, we propose that the CTOD can be used as a laboratory tool to characterize the nonlinear elastic parameters of soft materials by comparing results obtained from CTOD-based postprocessing and analytical results.
However, we note that these results are valid for zero-scale yielding only; in the presence of yielding, a hierarchy of scales must be considered. 

Because GNL constitutes the universal kinematic requirement for all finite-strain descriptions, the physical observations derived within the SVK framework establish a universally applicable baseline. We showed that nonlinear kinematics alone induce a strain-stiffening response in the crack, generating apparent crack-tip toughening (Fig. 7e) and a sharper, $\nu$-dependent CTOD (Figs. 2a and 6a). Real soft solids, however, exhibit diverse material  nonlinearities. For instance, the initial strain-softening typically observed in soft elastic solids actively counteracts this intrinsic GNL-driven stiffening, resulting in a blunted CTOD  \cite{Spagnoli2023}. We can definitively isolate and identify this constitutive softening mechanism only because the SVK baseline first establishes the opposing kinematic effect. Furthermore, dynamic fracture experiments on neo-Hookean hydrogels \cite{Livne2007, Wang2023} and numerical simulations utilizing the SVK material \cite{Lubomirsky2018, Pundir2024} both produce cracks propagating with oscillatory instability and supershear speeds. These observations highlight a fundamental question: what is the precise role of material and geometric nonlinearity across diverse fracture phenomena? From a material design perspective, do specific material nonlinearities suppress or enhance the effects of GNL? To answer this, we propose the SVK-based NLEFM framework as a foundational baseline. By first characterizing a fracture phenomenon using the SVK material, researchers can systematically attribute deviations from this behavior to specific material  nonlinearities. Future investigation in this direction involves the formulation of a generalized Mooney-Rivlin material model in which the SVK material constitutes a precise subspace within the parameter space. Continuously varying the parameters away from this SVK limit provides the exact mathematical route to systematically quantify the coupled effects of material nonlinearity against our universal kinematic baseline.

\section{Conclusion}\label{sect:conclusion}

In this work, we have elucidated the effects of GNL on the near-tip behavior of a static crack under mode I loading assuming plane-strain conditions. We proposed a CTOD-based post-processing to determine both the asymptotic behavior of the CTOD and a nonlinear length scale $\lengthSCd$ that defines the region in which GNL induces crack-tip behavior that deviates significantly from LEFM predictions. This framework permits us to establish a link between this solution and the far-field loading and geometric conditions. We derived analytic expressions describing the elastic crack-tip fields and defined the CTOD as a piecewise function with a transition at a distance $\lengthSCd$ from the crack tip. The proposed methodology and analytical results provide insight into the nonlinear behavior of a crack under mode I loading and offer a useful experimental tool for characterizing the nonlinear elastic parameters in soft materials. Furthermore, we have discussed the relationship between GNL and important phenomena such as supershear crack propagation speeds. We believe that the CTOD framework presented here can be extended to include material nonlinearity and serve as a foundation for the broader study of NLEFM.

\myacknowledgements{
The authors acknowledge the Swiss National Science Foundation for financial support under grant number 10003776. We thank Dr. Daniel Rayneau-Kirkhope for writing assistance.
}
\insertacknowledgements

\section*{CRediT authorship contribution statement}
\textbf{Raúl Lazo-Molina}: Formal analysis, Data curation, Investigation, Methodology, Software, Validation, Visualization, Writing -- original draft.\\
\textbf{Mokhtar Adda-Bedia}: Conceptualization, Supervision, Methodology, Formal Analysis, Investigation, Writing -- original draft.\\
\textbf{Mohit Pundir}: Methodology, Software, Supervision, Writing -- review \& editing.\\
\textbf{Rodrigo Arias}: Methodology, Formal analysis.\\
\textbf{David S. Kammer}: Conceptualization, Funding acquisition, Resources, Supervision, Writing -- review \& editing.

\section*{Declaration of competing interest}
The authors declare that they have no known competing financial interests or personal relationships that could have appeared to influence the work reported in this paper.

\section*{Code Availability}
The code used for the numerical simulations is available on \href{https://gitlab.ethz.ch/smec/papers-supp-info/2026/static-gnl-ctod}{ETH GitLab}.

\section*{Data Availability}
The data that support the findings of this study will be available in the final version of this manuscript.

\newpage

\appendix
\section{Numerical formulation}\label{sect:appendix_TL}

We consider a body in an undeformed reference configuration at pseudo-time $t = 0$. At time $t$, the equilibrium state of the body is given by a set of displacements ${}^t_0u_i$.  For a given time step $\Delta t$, displacements are updated by adding $\Delta u_i$ to the previous state ${}^t_0u_i$, resulting in an initial solution for the solution at $t+\Delta t$ given by ${}^{t+\Delta t}_0 u_i = {}^t_0u_i + \Delta u_i$. This initial solution, ${}^{t+\Delta t}_0 u_i$, is then refined through $n$-iterations of the Newton--Raphson method, leading to a converged numerical solution ${}^{t+\Delta t}_0 u_i^{(n)}$.

The incremental weak form of the problem considered here is conveniently defined in terms of the symmetric tensors $\secondPK$ and $\greenL$ \cite{Bathe1982}:
\begin{equation}
\int_{V} {}_0^{t+\Delta t}\secondPK[ij]\;\delta{}_0^{t+\Delta t}\!\greenL[ij]\;\mathrm{d}V
= {}^{t+\Delta t}\mathcal{R}\;,
\label{eqn:incremental_balance}
\end{equation}
where ${}^{t+\Delta t}\mathcal{R}$ contains all external forces applied at pseudo-time $t+\Delta t$, $\delta$ denotes a variational operator, and $V$ is the volume over which the integral is performed. 

After solving for ${}^{t+\Delta t}_0 u_i$, the incremental strain $\Delta \greenL$ is calculated considering its linear part $\Delta e_{ij}$ and its quadratic part $\Delta\eta_{ij}$, such that $\Delta \greenL=e_{ij}+\Delta\eta_{ij}$. These contributions to the strain are computed using \cref{eqn:greenL}, and $\Delta \greenL = {}^{t+\Delta t}\greenL - {}^{t}\greenL$:
\begin{eqnarray}
\Delta e_{ij}
&=& \frac{1}{2}\left(\Delta\gradU[i,j] + \Delta\gradU[j,i] 
   + {}^{t}_{0}\gradU[k,i]\,\Delta\gradU[k,j] 
   + \Delta\gradU[k,i]\,{}^{t}_{0}\gradU[k,j]\right)\;,
\label{eqn:strain_linear}\\
\Delta \eta_{ij}
&=& \frac{1}{2}\,\Delta\gradU[k,i]\,\Delta\gradU[k,j]\;.
\label{eqn:strain_nonlinear}
\end{eqnarray}
The incremental balance of mechanical energy thus takes the following final form:
\begin{equation}
\int_V \Delta\secondPK[ij]\;\delta \Delta\greenL[ij]\;\mathrm{d}V
\;+\;\int_V {}^t_0\secondPK[ij]\;\delta\Delta\eta_{ij}\;\mathrm{d}V
\;=\;
{}^{t+\Delta t}\mathcal{R}
\;-\;\int_V {}^t_0\secondPK[ij]\;\delta\Delta \greenL\;\mathrm{d}V\;,
\label{eqn:balance_energy}
\end{equation}
which is solved at every loading step considered in the loading procedure.

\section{CTOD-based procedure}

\begin{twocolfig}[!h]
    \centering
\includegraphics[width=1\linewidth]{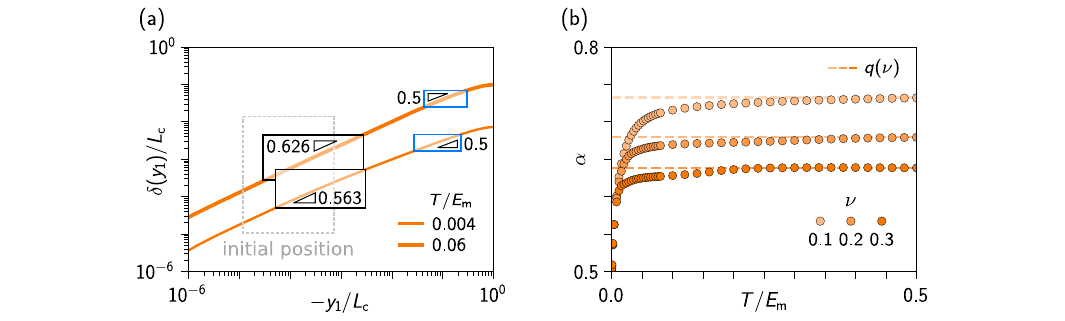}
    \caption{
    Computing the nonlinear asymptotic exponent. (a) The window of analysis used to track the evolution of the slope $\alpha$ as load on the system increases. Gray box shows the position of the analysis window in the undeformed configuration, while the black boxes shows the analysis window in the deformed configuration. Two curves show the same material ($\nu=0.3$) subject to two values of loading. Blue rectangles indicate the region of the CTOD where the slope (or exponent) is still $1/2$. (b) Evolution of the slope $\alpha$ with increasing load showing convergence to asymptotic exponent for different values of $\nu$.
}
    \label{fig:appendix_slopes}
\end{twocolfig}

\subsection{Finding the asymptotic solution}\label{sect:appendix_ctod_1}

Rather than tracking the exponent directly, we use a log-log scale to monitor the evolution of the slope denoted $\alpha$ near the crack tip as we increase the loading on the system, $T/\E$. To accurately obtain the slope close to the crack tip, while avoiding numerical noise, we select two specific nodes from the mesh that we use to define an analysis window, from which we compute $\alpha$ (the width and position of this window are indicated by the gray box in \cref{fig:appendix_slopes}a). As loading on the system is increased, the window moves with the selected nodes and the slope converges to a fixed value, see \cref{fig:appendix_slopes}b. We finally average these converged $\alpha$-values to obtain $q$. The values of $q$ obtained using this procedure are shown in \cref{fig:slopes}b.

\subsection{Finding the nonlinear length scale}\label{sect:appendix_ctod_2}

We have defined $\lengthSCd$ as the horizontal coordinate of the crossover between the inner solution $\delta^\text{inner}$ and outer solution $\delta^\text{outer}$, given by  $\delta^\text{inner}(\lengthSCd)=\delta^\text{outer}(\lengthSCd)$, see \cref{fig:lambda_nl_cur}a. For a given value of $\nu$, we take the appropriate value of $q$ (\cref{fig:slopes}b) and we fit the CTOD data to the power laws $\delta^\text{inner}=\tilde{a} \cdot (-y_1)^{q(\nu)}$ and  $\delta^\text{outer}=\tilde{b} \cdot (-y_1)^{1/2}$, where $\tilde{a}$ and $\tilde{b}$ are considered as unknown positive scalar coefficients. We then find the intersection of these curves and thus obtain the nonlinear length scale as $-y_1=\lengthSCd$.

We facilitate this procedure by normalizing the CTOD $\delta$ by $(-y_1)^{1/2}$ on a log-log scale (see \cref{fig:appendix_lambda_nl_cur}b). In this representation, the outer region appears as a horizontal line, while the inner region presents a slope of $q(\nu) - 1/2$. Since LEFM directly relates $\tilde{b}$ to $K_I$, the obtained solution allows us to link the nonlinear asymptotic region with the far-field conditions. We therefore use $\tilde{b}$ to compute $K_I^{(\delta)}$, where the superscript indicates that we obtained the value of $K_I$ directly from the CTOD.

\begin{twocolfig}[!h]
    \centering
\includegraphics[width=1.0\linewidth]{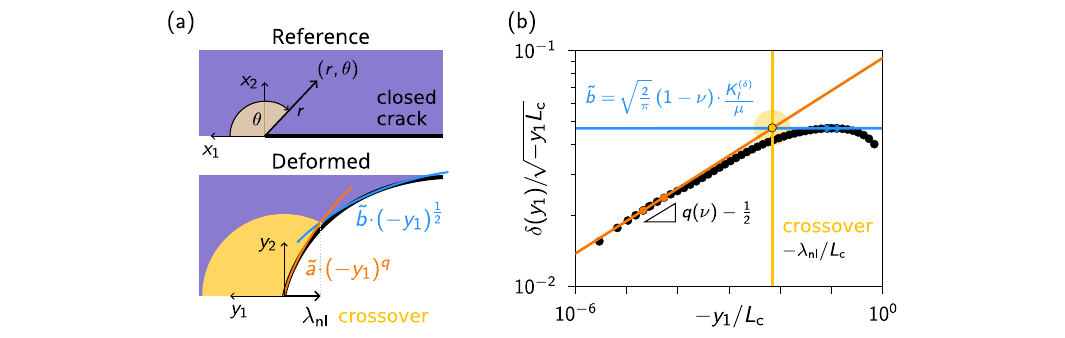}
    \caption{Computation of the nonlinear length scale $\lengthSCd$ in the deformed configuration. (a) Schematic of the two solutions and the fitting procedure to compute $\lengthSCd$, showing the reference configuration and the deformed configuration above and below, respectively. (b) The fitting is undertaken by considering the CTOD $\delta\!\left(y_1\right)$ normalized by $\left(-y_1\right)^{1/2}$, such that the horizontal blue line indicates the region where the exponent is $1/2$. The expression in blue corresponds to LEFM solution for the coefficient $\tilde{b}$ of the assumed outer solution for the mode I plane strain case ($T/\E=0.02$ and $\nu=0.3$). The orange curve shows the fitting using the value of $q$ obtained according to \cref{sect:appendix_ctod_1}
}
    \label{fig:appendix_lambda_nl_cur}
\end{twocolfig}

\subsection{A pseudo-reference description}\label{sect:appendix_sref}

The LEFM framework is defined with respect to a pseudo-reference configuration in which horizontal displacements of the crack surface are assumed to be zero. Here, we consider the CTOD profile setting all horizontal displacements to zero. In this configuration, we consider the  CTOD $\delta$ to be given by the pair $\left(-x_1,\delta\!\left(x_1 \right) \right)$, where $-L_\mathrm{c}\leq x_1 \leq 0$ and $\delta\!\left(x_1\right)=y_2$, see \cref{fig:appendix_lambda_nl_ref}a. Following the same procedure as presented in \cref{subsect:asymptotic_behavior,sect:appendix_ctod_1}, the asymptotic exponent in this configuration is found to be $3/4$, that is, $y_2\sim \left(-x_1\right)^{3/4}$, and is independent of $\nu$. We plot the relationship between $\lengthSCr$ and $K_I/\mu$ in \cref{fig:appendix_lambda_nl_ref}c and its inset. 

\begin{twocolfig}[htbp!]
    \centering
\includegraphics[width=1.0\linewidth]{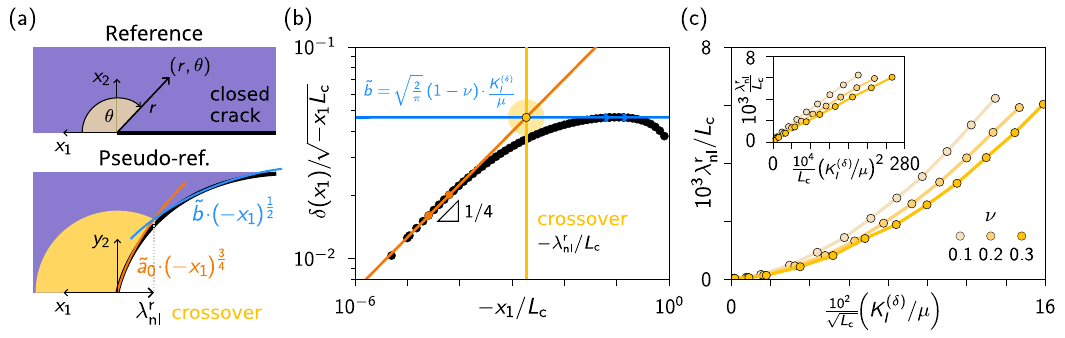}
    \caption{Computation of the nonlinear length scale $\lengthSCr$ in the pseudo-reference configuration. (a) Schematic of the two solutions and the fitting procedure to compute $\lengthSCr$, showing the reference configuration and the pseudo-reference configuration above and below, respectively. (b) The fitting is undertaken considering the CTOD $\delta\!\left(x_1\right)$ normalized by $\left(-x_1\right)^{1/2}$; the slope of the curve in the inner region is constant for all values of $\nu$. (c) $\lengthSCr/L_\mathrm{c}$ as a function of normalized stress intensity factor computed from CTOD $K_I^{(\delta)}/\mu$. The inset shows the scaling of $\lengthSCr/L_\mathrm{c}$ with respect to $\left(K_I^{(\delta)}/\mu\right)^2$.
}
    \label{fig:appendix_lambda_nl_ref}
\end{twocolfig}

\section{Derivations of asymptotic analysis}\label{sect:appendix_derivations}

\subsection{Asymptotic expressions}\label{sect:appendix_tensors}

We use the assumed asymptotic solution defined in \cref{eqn:sol_form} to obtain the kinematic quantities required for the analysis in \cref{subsect:inner_sol}. Considering polar coordinates, we obtain the components $\defGrad[ij]=\pd y_i/\pd x_i$ as:
\begin{align}
\defGrad[i1] &= a_i^{1-m_i}\,r^{m_i-1}(m_i U_i\cos \theta - \dot{U}_i\sin \theta) \;, \label{eqn:Fa1} \\
\defGrad[i2] &= a_i^{1-m_i}\,r^{m_i-1}(m_i U_i\sin \theta +\dot{U}_i\cos \theta )\;. \label{eqn:Fa2}
\end{align}
We use the conditions in \cref{eqn:conds} to examine the asymptotic behavior of the quantities $\firstI$ and $\jacDet$ as $r\to 0$; we find that:
\begin{align}
\firstI  &\simeq a_2^{2(1-m_2)}\,r^{2(m_2-1)} \angOne (\theta)\; , \label{eqn:Ia}\\
\jacDet &\simeq a_1^{1-m_1}a_2^{1-m_2}\,r^{m_1+m_2-2}\angTwo(\theta) \; , \label{eqn:Ja}
\end{align}
where the angular functions $\angOne(\theta)$ and $\angTwo(\theta)$ are given by:
\begin{eqnarray}
\angOne(\theta) &=& m_2^2U_2^2+\dot{U}_2^2\;,\\
\angTwo(\theta) &=& m_1 \angSoll[1] \dot{U}_2-m_2 U_2 \dot{U}_1\;.
\end{eqnarray}

We can then use the asymptotic variables defined in \crefrange{eqn:Fa1}{eqn:Ja} to determine the dominant term controlling the singular behavior of $\energyDen$ and $\firstPK[ij]$  as $r\to 0$. In the case of $\energyDen$ (\cref{eqn:svk_energy_den_IJ}), given the conditions that the exponents $m_i$ must satisfy (\cref{eqn:conds}), we see that $I_1^2$ diverges faster than both $I_1$ and $J$. Hence, $\energyDen$ is dominated by $\firstI^2$, such that:
\begin{equation}
    \energyDen \simeq \frac{\kappa\mu}{4} a_2^{4(1-m_2)}\,r^{4(m_2-1)}M^2(\theta)\;,
\end{equation}
and therefore:
\begin{equation}
    \energyDen \propto r^{4(m_2-1)}\;.
    \label{eqn:energy_den_scaling}
\end{equation}
Similarly, it can be shown that the stress components $\firstPK[ij]$  in \cref{eqn:pk1_alternative} are asymptotically dominated by the following terms:
\begin{eqnarray}
P_{11} &\simeq&
\mu a_1^{1-m_1}a_2^{2(1-m_2)}r^{m_1+2m_2-3}
\left[
\kappa M(\theta)\left(m_1 U_1\cos \theta -\dot{U}_1\sin \theta\right)
-
N(\theta)\left(m_2 U_2\sin \theta +\dot{U}_2\cos \theta\right)
\right] 
\nonumber\\
&\simeq& \kappa\mu IF_{11} - \mu JF_{22} 
\;,\\
P_{12} &\simeq& \mu a_1^{1-m_1}a_2^{2(1-m_2)}r^{m_1+2m_2-3}
\left[\kappa M(\theta)\left(m_1 U_1\sin \theta + \dot{U}_1\cos \theta\right)
+ N(\theta)\left(m_2 U_2\cos \theta - \dot{U}_2\sin \theta\right)
\right]
\nonumber\\
&\simeq& \kappa\mu IF_{12} + \mu JF_{21} 
\;,\\
P_{21} &\simeq& \kappa\mu a_2^{3(1-m_2)}r^{3(m_2-1)}M(\theta)\left(m_2 U_2\cos \theta - \dot{U}_2\sin \theta\right)
\nonumber\\
&\simeq& \kappa\mu IF_{21}
\;,\\
P_{22} &\simeq& \kappa\mu a_2^{3(1-m_2)}r^{3(m_2-1)}M(\theta)\left(m_2 U_2\sin \theta +\dot{U}_2\cos \theta\right)
\nonumber\\
&\simeq& \kappa\mu IF_{22}
\;.
\end{eqnarray}
Their scaling is as follows:
\begin{eqnarray}
\firstPK[11], \firstPK[12] &\propto& r^{m_1+2m_2-3}\;,
\label{eqn:pk1_scaling_1i}\\
\firstPK[21], \firstPK[22] &\propto& r^{3(m_2-1)} \;,
\label{eqn:pk1_scaling_2i}
\end{eqnarray}
where from \cref{eqn:conds}, we have that $m_1+2m_2-3 > 3(m_2-1)$ and $3(m_2-1)<0$, so that the dominant behavior are those of $\firstPK[21]$ and $\firstPK[22]$.

\subsection{Balance equation for $i=2$}\label{subsect:appendix_eqU2}

Here, we consider how to obtain a solution for  $\angSoll[2]\left(\theta\right)$. Considering the behavior at divergence of the components $\firstPK[2j]\propto \firstI \defGrad[2j]$ (see \cref{sect:appendix_tensors}), we find that the approximate form of the equilibrium equation $\firstPK[2 j,j]=0$ (\cref{eqn:BVP}) is:
\begin{equation}
(\firstI \defGrad[2j])_{,j}=0 \; .    
\label{eqn:eqf0}
\end{equation}
We then substitute the definition given in \cref{eqn:Fa2,eqn:Ia} into \cref{eqn:eqf0} to find the final balance equation that needs to be solved:
\begin{equation}
\left(\ddot{U}_2+m_2^2\angSoll[2]\right)\angOne+2m_2\left(m_2-1\right)\angSoll[2]\,\angOne+\dot{U}_2\dot{\angOne} =0 \; .
\end{equation}

\subsection{Balance equation for $i=1$}\label{subsect:appendix_eqU1}

We consider now the equilibrium equation $\firstPK[1 j,j]=0$ (\cref{eqn:BVP}), which, in view of \cref{eqn:pk1_alternative}, can be expressed as:
\begin{equation}
\kappa (\firstI \defGrad[1j])_{,j}-
\epsilon_{j l}(\jacDet \defGrad[2l])_{,j}=0 \;.
    \label{eqn:eqv0}
\end{equation}
By considering the asymptotic forms defined in \crefrange{eqn:Fa1}{eqn:Ja}, we determine that the dominant term in the above expression is proportional to $r^{2m_2+m_1-4}$.
Using the latter, together with \cref{eqn:Fa2,eqn:Ia}, and given the solutions of $m_2$ (\cref{eqn:m_2}) and $U_2$ (\cref{eqn:U2}), we find that \cref{eqn:eqv0} leads to the following approximate equation for $m_1$ and $\angSoll[1]$:
\begin{equation}
\kappa \left( \left(m_1+2m_2-2\right)m_1U_1\angOne+\dot{U}_1\,\dot{\angOne}+\ddot{U}_1\,\angOne\right)=
 (m_1+m_2-2)\dot{U}_2\,\angTwo-m_2U_2 \,\dot{\angTwo}\;\; .
\end{equation}

\newpage
\bibliographystyle{\mybibliostyle}
%

 %

%

\end{document}